# Impedance and electrical properties of Cu doped ZnO thin films

[1]P. Samarasekara, [1]Udumbara Wijesinghe and [2]E.N. Jayaweera
[1]Department of Physics, University of Peradeniya, Peradeniya, Sri Lanka
[2]Department of Chemistry, University of Peradeniya, Peradeniya, Sri Lanka

**Abstract**
Cu doped transparent ZnO thin films were spin coated on conductive glass substrates. The samples were subsequently annealed in air for 1 hour at 500 $^0$C in order to form the phase of ZnO. ZnO samples were doped with different Cu molar percentages up to 5%. The impedance and photovoltaic properties of sample were measured. Photocurrent and photovoltage of doped and undoped samples were measured in $KI/I_2$ electrolyte. Adding a trace amount of Cu improved the conducting properties of ZnO samples without changing other basic properties of ZnO. The photocurrent gradually increases with the doping concentration due to the high conducting properties of Cu. Investigation was carried out only up to the doping concentration of 5%, because higher doping concentrations may significantly influence the other properties of ZnO such as transparence of the film. Impedance of samples was determined by fitting the data to an equivalent circuit. The impedance reaches the maximum value at Cu concentrations of 3%.

## 1. Introduction:

ZnO finds potential applications in electronic devices, optoelectronic devices, acoustic wave devices, piezoelectric devices, photocells [1] and gas sensors. Owing to the higher band gap of ZnO, it can absorb the ultra violet wave range in the spectrum and it is transparent to visible light. Thin films of undoped ZnO have been previously deposited using the reactive dc sputtering by us [1]. Optical properties of sol-gel-processed undoped ZnO films have been investigated [2]. In addition, structural and optical properties of undoped ZnO films synthesized using dc and rf sputtering have been studied [3]. ZnO films have been fabricated on Si (1 1 1) substrates using pulsed laser deposition technique [4]. Furthermore, transparent conducting cobalt doped ZnO films have been grown [5]. Li-, P- and N-doped ZnO thin films have been prepared by pulsed laser deposition [6]. Structural and optical properties of ZnO thin films synthesized on (111) $CaF_2$ substrates using magnetron sputtering have been investigated [7]. Effect of deposition temperature on the crystallinity of Al-doped ZnO thin films grown on glass substrates by RF magnetron sputtering method has been studied [8]. Previously carbon nanotubes have been synthesized by us [12]. Energy gap of semiconductor particles doped with salts were determined by us [13]. Copper oxide was fabricated using reactive dc sputtering by us [14]. Low Cost p-$Cu_2$O/n-CuO Junction was prepared using oxidization [15]. ZnO indicates some magnetic properties. Magnetic properties were theoretically explained by us [16, 17, 18, 19, 20, 21].

In this report, the impedance and photovoltaic properties of Cu doped ZnO films synthesized using spin coating method have been described. The spin coating technique is inexpensive compared to deposition technique incorporated with vacuum systems. Initially, samples were prepared at different rpm values of spin coating system for different time durations to find the best deposition conditions. All the films grown at the best deposition conditions were subsequently annealed at different temperatures for different durations in air. The variation of electrical properties of Cu doped ZnO samples with the doping concentrations was investigated.

A chemical method was used to form the ZnO phase in the thin film. The formation of ZnO phase in thin film from has been confirmed using energy gap calculations as described in one of our other research articles.

## 2. Experimental:

ZnO sols were prepared by using zinc acetate dihydrate ($Zn(CH_3COO)_2\cdot 2H_2O$), anhydrous ethanol and monoethanolamine (MEA) as the solute, solvent and sol stabilizer, respectively. Zinc acetate was first grinded and dissolved in a mixture of ethanol and monoethanolamine at room temperature. The molar ratio of MEA to zinc acetate was kept at 1:1. Five doped sols were prepared by adding copper acetate dihydrate ($Cu(CH_3COO)_2\cdot 2H_2O$) to the mixture with molar percentage of Cu varying from 1% to 5%. The resulting solutions were stirred by a magnetic stirring apparatus at 70 °C for an hour. At last, transparent ZnO sols were formed. In the sol, the Zn concentration was 0.5 mol/L. The prepared sols were kept for 24 hours at room temperature. Then the thin films were prepared by the spin-coating method on conductive glass (ITO) substrates which had been pre-cleaned by detergent, and then cleaned in methanol and acetone for 10 min each by using ultrasonic cleaner and then cleaned with deionized water and dried. The spin-coating time was 30 s and the spin-speed was 2000 rpm. After coating, the samples were first dried at 200 °C for 10min, and then were annealed at 500 °C in ambient atmosphere for an hour. The films prepared by doped sols of Cu by molar percentage varying from 0% to 5% were labeled as samples A–F respectively.

The ionic conductivity of the samples were determined by the AC complex impedance method with a computer controlled Solatron SI–1260 impedance analyzer in the frequency range 20 Hz –10 MHz. The temperature of the sample was kept at 25 $^0$C for 30 min prior to the measurement. The ionic conductivity of each sample was extracted from the corresponding impedance plots and this procedure repeated for all the six compositions. The photocell was prepared using doped or undoped ZnO film, platinum electrode and $KI/I_2$ liquid electrolyte. (I–V) characteristics of the cells were measured under the dark condition and illumination of 1000 mW cm$^{-2}$ (AM 1.5) simulated sunlight using a homemade computer controlled setup coupled to a Keithley 2000 multimeter and a potentiostat/galvonostat HA-301. A Xenon 500 lamp was used with AM 1.5 filter in order to obtain the simulated sunlight with above intensity.

## 3. Results and Discussion:

The impedance behavior of the Cu doped and un-doped ZnO thin films under room temperature is shown in figure 1. The figure shows the relationship between the real and imaginary parts of the complex impedance. Evidently the graphs contain a semi-circle and a straight line portions. The semi-circle part elongates as the molar concentration of Cu increases from 0% to 3% and shrinks at 4%. The curves of 1% and 2% overlap each other. The slope of the straight line part is approximately equals to 0.3.

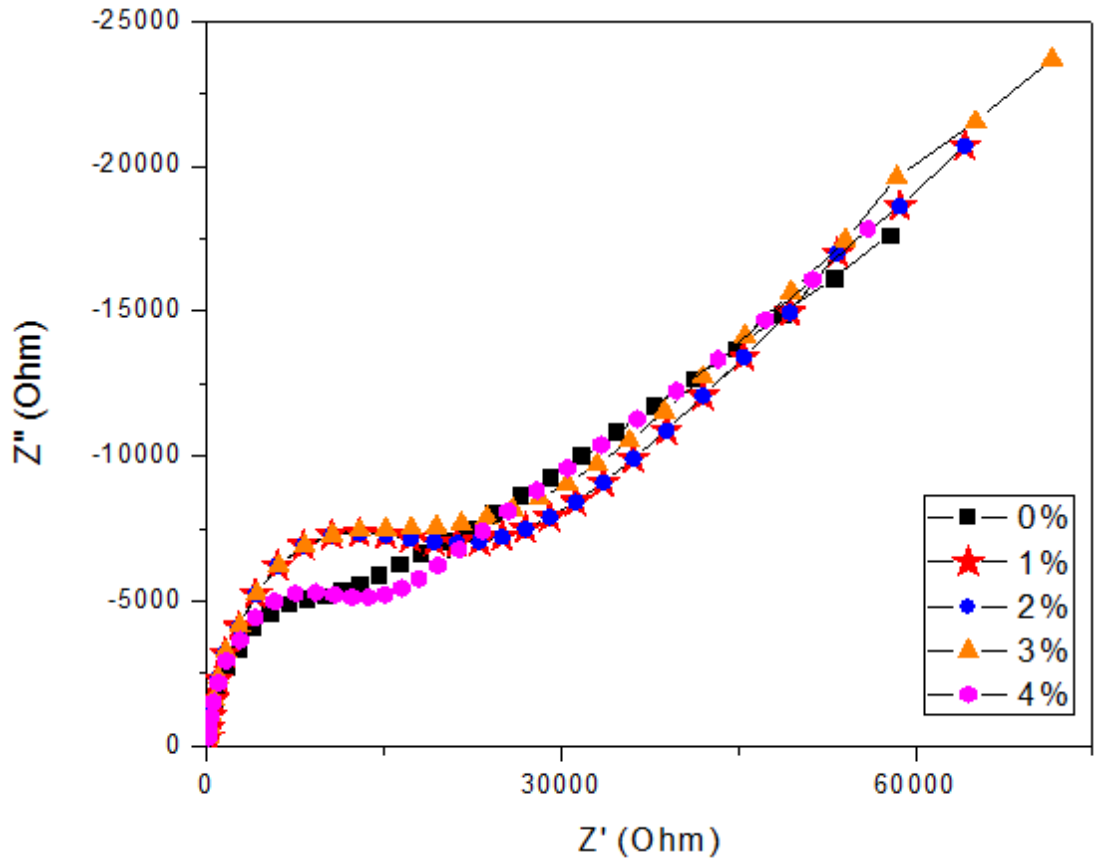

Figure 1: The impedance plots of Cu doped and undoped ZnO thin films.

A standard model in the Gamry EIS300 electrochemical impedance spectroscopy software was used to interpret the data and it revealed a circuit that contains two resistors capacitor and Warburg impedance as shown in figure 2.

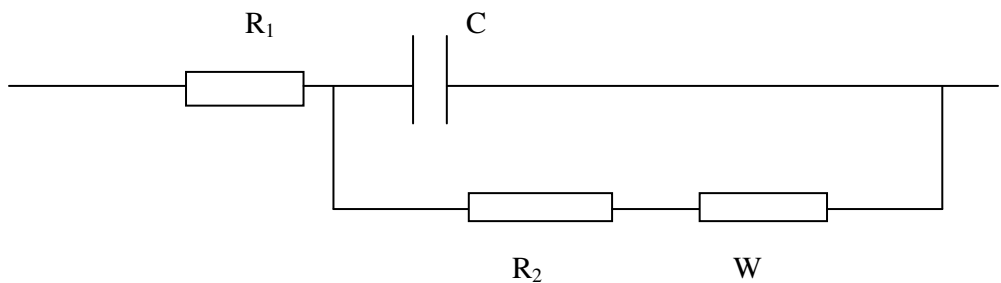

Figure 2: Equivalent circuit corresponding to figure 1.

The values of $R_1$ and $R_2$ of this equivalent circuit were estimated manually and the estimated values of the components of the equivalent circuit are tabulated in table 1.

| Concentration % | $R_1/\Omega$ | $R_2/\Omega$ |
|---|---|---|
| 0 | 102.02 | 21315.71 |
| 1 | 102.56 | 29244.13 |
| 2 | 102.56 | 29244.13 |
| 3 | 111.26 | 29234.89 |
| 4 | 108.56 | 25047.69 |

Table 1: Corresponding values $R_1$ and $R_2$ for the equivalent circuit given in figure 2.

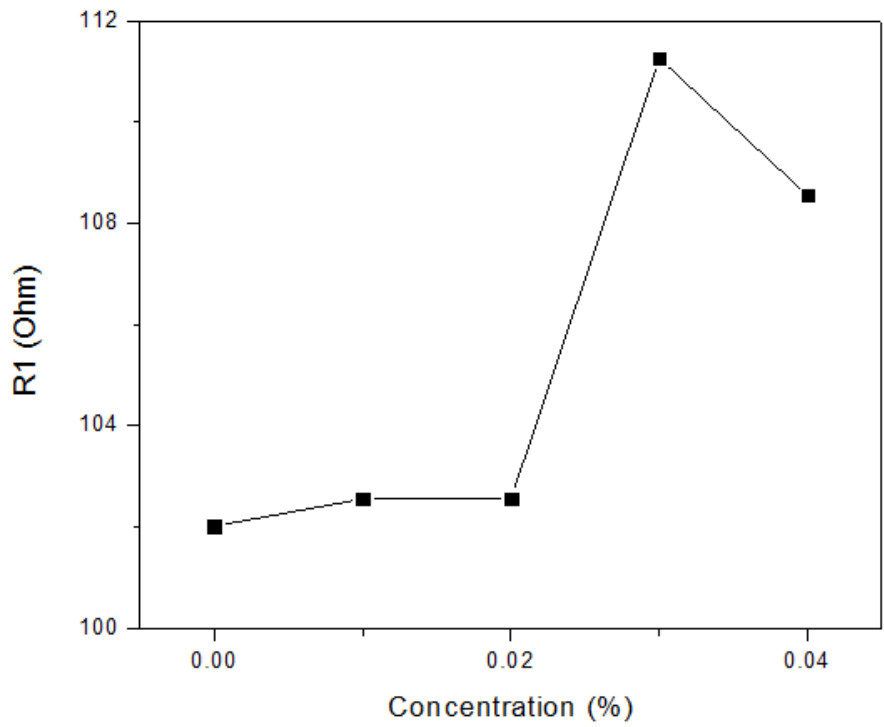
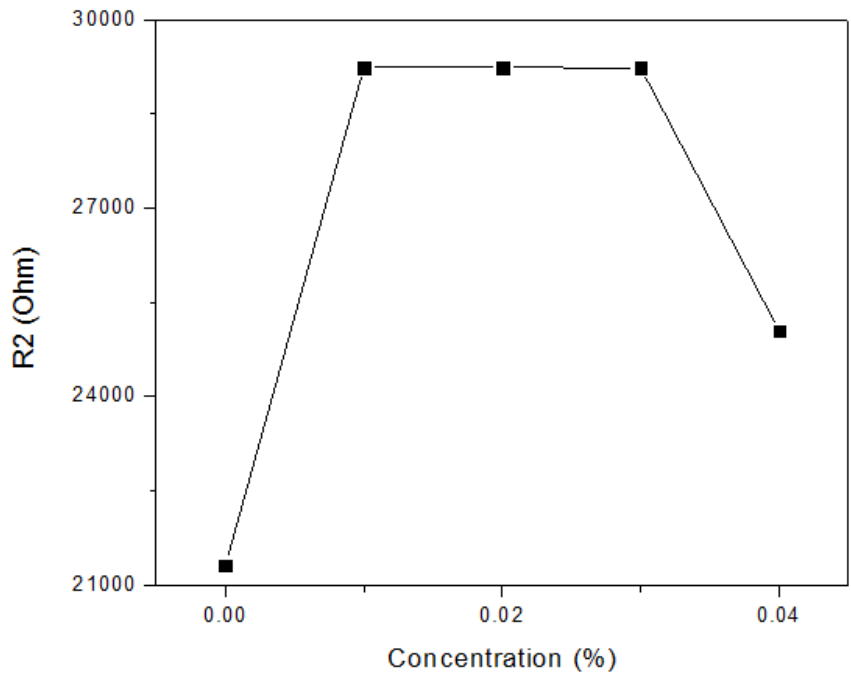

Figure 3: The variation of $R_1$ and $R_2$ as a function of the doping concentration of Cu

The electrical resistance as a function of doping concentration in the starting solution shows an increase with the copper concentration, reaching a maximum value at certain ratio (3% in both cases), and then decrease in resistance values is observed when the copper concentration increases to 4%. In the case of pure ZnO, the important factor which effects on the mobility is the presence of grain boundaries. Grain boundaries primarily act as scattering sites and potential barriers leading to reduce the carrier mobility by increasing the resistance. For low concentration of Cu, the increase of resistance is due to the increase of Cu atoms that are incorporated in to the ZnO lattice in the Zn sites by supplying a hole for each Cu atom until the maximum solubility of Cu in to the ZnO lattice reaches (the maximum values at resistance curves). This indicates that the Cu ions introduce in Zn lattice play the role of an acceptor type impurity. For higher Cu concentrations, segregation of Cu takes place in the grain boundaries or interstices causing a fall in resistance [9].

The photocell was prepared using doped or undoped ZnO film, platinum electrode and $KI/I_2$ liquid electrolyte. I-V characteristic curves of undoped ZnO thin film sample is shown in figure 4. The area of all the samples is 1 $cm^2$. The photocurrent (or photo-voltage) has been defined as the difference between light and dark currents (or voltages). The photocurrent at V = 0 is 0.038 mA, and the photo-voltage at I = 0 is V = 0.488 V. Here dashed and solid lines indicate the dark and light I-V curves, respectively. The calculated efficiency corresponding to this photocurrent and photo-voltage is 0.07%. Figure

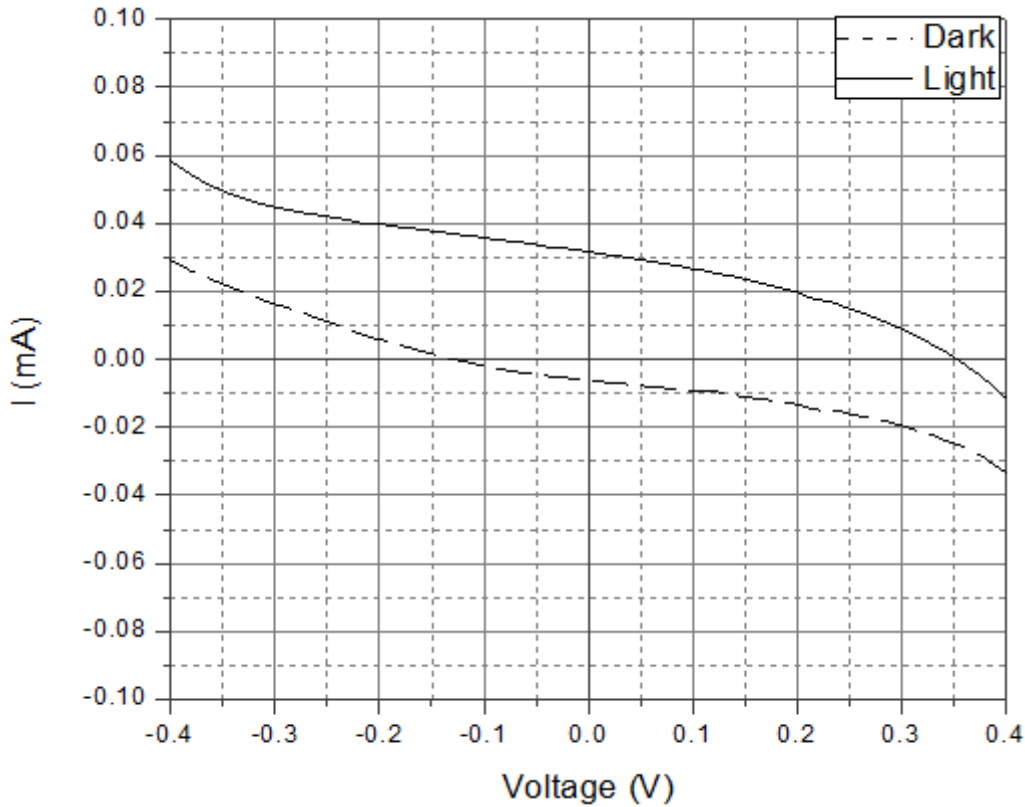

Figure 4: I-V curve of pure ZnO thin film device under dark and light illumination conditions.

The photocurrent and photo-voltage vary with the doping concentration of Cu. The calculated photo current ($I_P$), photo voltage ($V_P$), output power (Pout) and the efficiencies ($\eta$) for doped samples are shown in the table 2. According to figures, photo current increases with the doping concentration while photo voltage decreases. A significant efficiency of 0.13% can be seen only in 5% while other doping concentrations result an average efficiency of 0.07%. The trace amount of copper added can be the reason for the slight increase of efficiency.

| Concentration % | $V_p$ (V) | $I_p$ (mA) | $P_{out}$ (mW) | $\eta$ (%) |
|---|---|---|---|---|
| 0 | 0.488 | 0.038 | 0.018544 | 0.07 |
| 1 | 0.369 | 0.046 | 0.016974 | 0.07 |
| 2 | 0.330 | 0.047 | 0.015510 | 0.06 |
| 3 | 0.294 | 0.066 | 0.019404 | 0.08 |
| 4 | 0.255 | 0.069 | 0.017595 | 0.07 |
| 5 | 0.243 | 0.138 | 0.033534 | 0.13 |

Table 2: Vp, IP, Pout and $\eta$ for Pin = 25×10-3 w at different doping concentrations of Cu

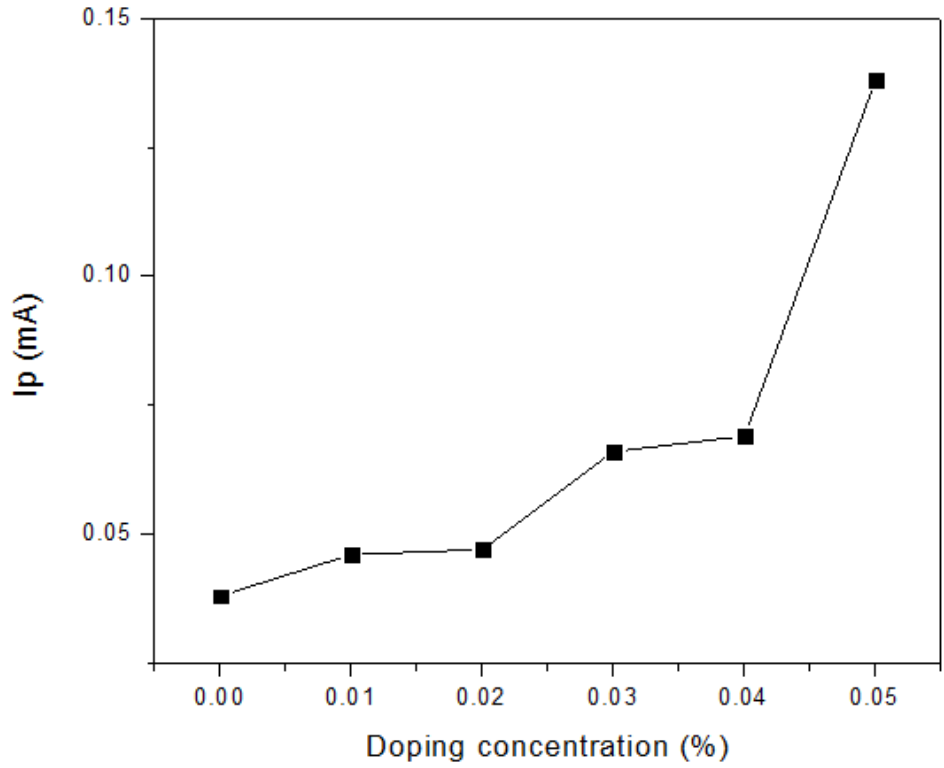

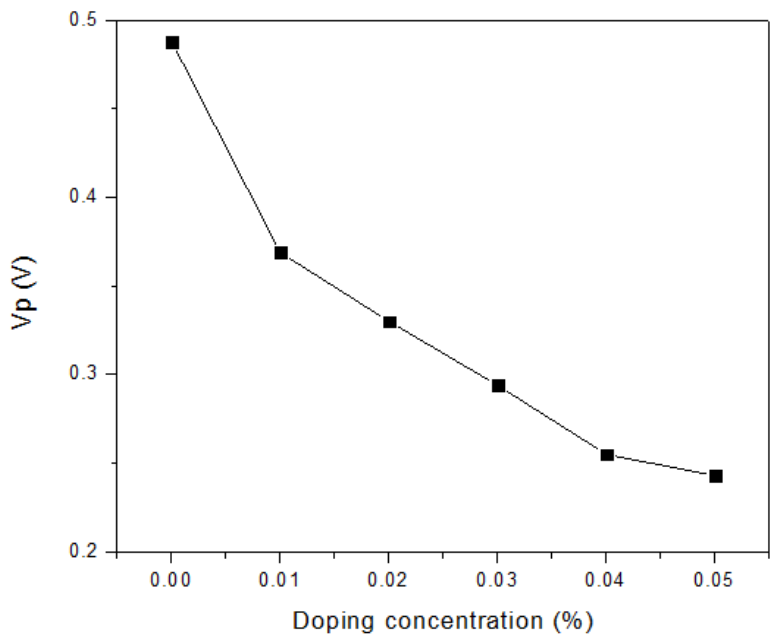

Figure 5: Photo current (Ip) and Photo voltage (Vp) as a function of doping concentration

The reduction of optical band gap in doped ZnO thin films at room temperature may be attributed to the sp-d exchange interaction between the band electrons and the localized d electron of the $Cu^{2+}$ ion substituting $Zn^{2+}$ ions [10, 11]. The s-d and p-d exchange give rise to negative and positive corrections to the conduction band and the valence band, respectively, leading to the band gap narrowing generation of photon hole pair become efficient, assuming all incident photons with energy equal or greater than energy gap will be absorbed generating e-h pairs, and hence the photo current increases. On the other hand, the photo voltage decreases due to the increase of the band gap. The AFM images of 5% Cu doped and undoped ZnO samples are shown in figures 6 and 7, respectively. These images imply that the film samples are fairly uniform and free of voids.

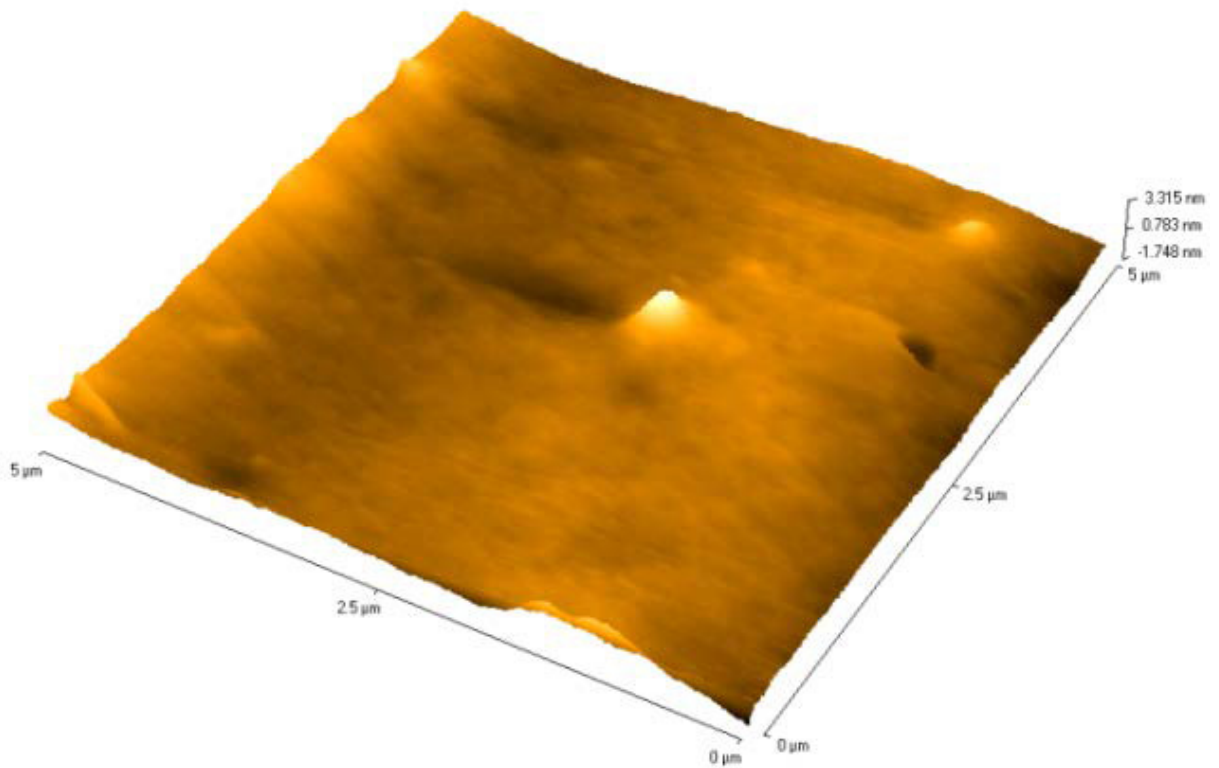

Figure 6: AFM image of 5% Cu doped ZnO film.

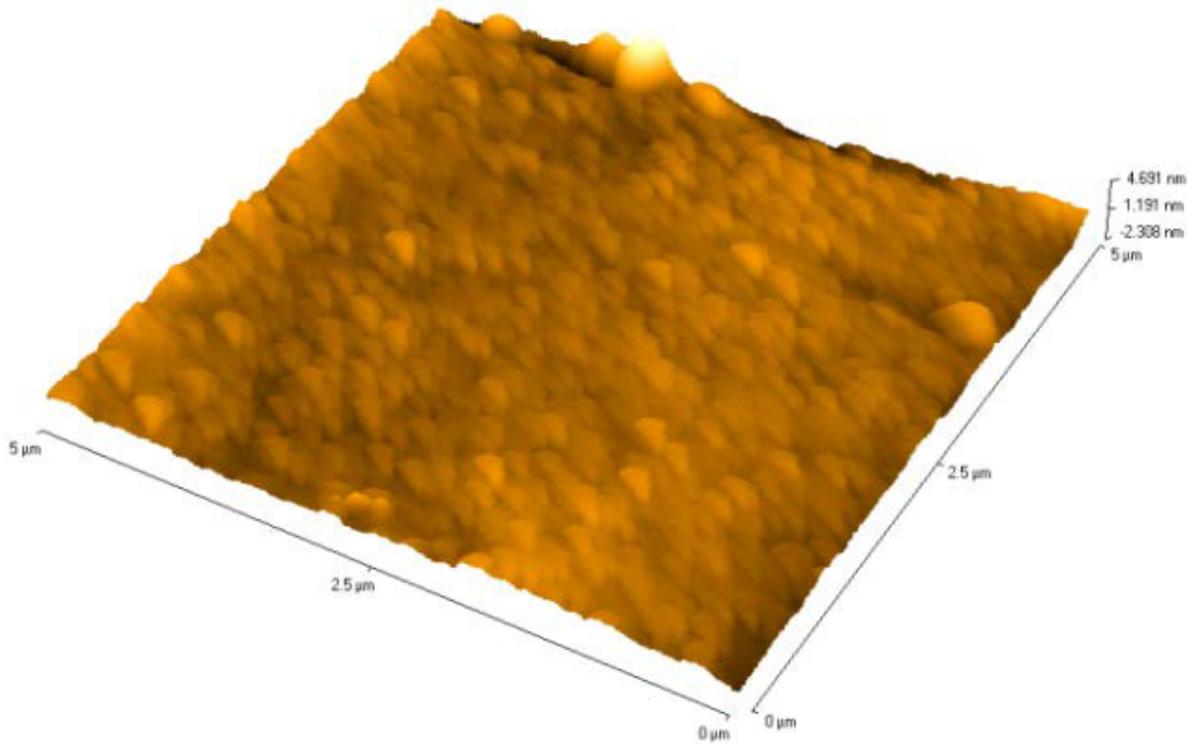

Figure 7: AFM image of undoped ZnO film.

## 4. Conclusion:

ZnO phase in thin film could be synthesized using low cost spin coating technique. The electrical resistances of equivalent circuit corresponding to our thin film samples reach a maximum at 3% of Cu doping concentration. At lower doping concentrations such as 3%, grain boundaries primarily act as scattering sites and potential barriers leading to reduced carrier mobility by increasing the resistance. At higher doping concentrations, the conductivity enhances by the energy levels added by the impurity. As a result, the resistance decreases at higher doping concentrations. The photocurrent gradually increases with doping concentration due to the adding of impurity energy levels. As a result, the photo-voltage gradually decreases with doping concentration. It is obvious that higher photocurrents can't be obtained for ZnO due to the higher band gap of ZnO. ZnO is known as a material with higher resistivity. The conductivity of ZnO could be slightly enhanced by doping a trace amount of Cu without altering other properties of ZnO. The enhancement of conductivity of ZnO is important in the applications of photocells. AFM images indicate that our ZnO films are uniform.